\documentclass[aps,prb,twocolumn,prabib,amsmath,amssymb,superscriptaddress,notitlepage]{revtex4-1}
\usepackage{graphics}
\usepackage{bm}
\usepackage{dcolumn}
\usepackage{natbib}
\usepackage{epstopdf}
\usepackage{float}
\usepackage{graphicx}
\usepackage{epsfig}
\usepackage[pdfstartview=FitH]{hyperref}
\usepackage{color}
\usepackage{appendix}
\usepackage{ulem}

\newcommand{\kk}{{\bf k}}

\newcommand{\rr}{{\bf r}}

\begin{document}
\title{Nature of continuous phase transitions in interacting topological insulators}
\author{Tian-Sheng Zeng}
\affiliation{Department of Physics and Astronomy, California State University, Northridge, California 91330, USA}
\author{W. Zhu}
\affiliation{Theoretical Division, T-4 and CNLS, Los Alamos National Laboratory, Los Alamos, New Mexico 87545, USA}
\author{Jian-Xin Zhu}
\affiliation{Theoretical Division, T-4 and CNLS, Los Alamos National Laboratory, Los Alamos, New Mexico 87545, USA}
\affiliation{Center for Integrated Nanotechnologies, Los Alamos National Laboratory, Los Alamos, New Mexico 87545, USA}
\author{D. N. Sheng}
\affiliation{Department of Physics and Astronomy, California State University, Northridge, California 91330, USA}
\date{\today}
\begin{abstract}
We revisit the effects of the Hubbard repulsion on quantum spin Hall effects (QSHE) in two-dimensional quantum lattice models. We present both unbiased exact diagonalization and density-matrix renormalization group simulations with numerical evidences for a continuous quantum phase transition (CQPT) separating QSHE from the topologically trivial antiferromagnetic phase.
Our numerical results suggest that, the nature of CQPT exhibits distinct finite-size scaling behaviors, which may be consistent with either Ising or XY universality classes for different time-reversal symmetric QSHE systems.
\end{abstract}
\maketitle
\section{Introduction}
Landau continuous phase transitions classified by spontaneous symmetry breaking of local order parameters~\cite{Landau1937} are important concepts in condensed matter physics that lie at the heart of our understanding of various aspects such as quantum magnetism and superconductivity. However topological phase transitions among topological phases of matter which are generally indistinguishable by any local order parameters~\cite{Wen1990}, should require a change in topological invariant.  For a symmetry-protected topological phase, its topological characterization  is only well-defined in the presence of symmetry, like fermionic QSHE with time-reversal symmetry can be characterized by a $Z_2$ topological index, which is the main concern of our paper. In the interacting systems with topologically non-trivial structure, the interplay between topology, symmetry and interaction, may lead to complex nature for the quantum phase transition, possibly a first-order transition~\cite{Varney2011,Amaricci2015,Roy2016}. Nevertheless, whether or not a continuous phase transition can be accompanied by a change of topological order, is an intricate open question~\cite{Samkharadze2016,Lee2015,Castelnovo2008,Hamma2008}, which motivates us to reinvestigate the strong correlation effects on the interacting topological insulators with $Z_2$ topological index.

Recent studies on topological insulators have indicated such a concrete example of interaction-driven continuous quantum phase transition (CQPT) from $Z_2$ topological order to antiferromagnetism, where the universal continuous evolutions of physical quantities are expected~\cite{Ran2006,Moon2012,Whitsitt2016}. Within the Kane-Mele-Hubbard (KMH) model~\cite{Kane2005a,Kane2005b}, Xu and Moore proposed a CQPT from the QSHE to the trivial Mott insulator driven by interactions~\cite{Xu2006}. In the strong coupling limit, Rachel and Le Hur first derived its effective spin Hamiltonian up to second order perturbation, and concluded that the Mott antiferromagnetism (AFM) is in the transverse $xy$-plane, instead of in the longitudinal $z$-direction~\cite{Rachel2010}. And this scenario including the CQPT from the QSHE to the trivial $xy$-AFM was supported by numerical studies including quantum Monte Carlo (QMC) methods~\cite{Hohenadler2011,Zheng2011,Hohenadler2012,Assaad2013a,Assaad2013b,Hohenadler2014}, the variational cluster approach~\cite{Yu2011,Budich2012}, and the mean field theory~\cite{Pesin2010,Murakami2007,Rachel2010,Vaezi2012,Chen2015,Fiete2012,Reuther2012,Wu2012,Liu2013,Laubach2014}. In QMC simulations of spin order, a finite size analysis shows that the transverse long-range spin correlation $\langle S_{\rr}^xS_{\rr'}^x\rangle$ remains a robust finite value as the distance $|\rr-\rr'|$ increases in the antiferromagnetic regime, while the longitudinal long-range spin correlation $\langle S_{\rr}^zS_{\rr'}^z\rangle$ vanishes as the system size increases~\cite{Zheng2011,Hohenadler2012,Assaad2013a,Hohenadler2014}. In Ref.~\cite{Assaad2013a}, a Curie-law signature in the magnetic susceptibility is identified by adiabatically inserting a $\pi$ flux. Early studies based on mean-field theory~\cite{Pesin2010,Murakami2007,Rachel2010,Vaezi2012,Chen2015} predicted the existence of intermediate topological antiferromagnetic phases at certain moderate Hubbard repulsion, making the nature of CQPT more {\it intricate}. However, there is no signal of an intermediate topological phase
being detected by recent numerical QMC simulations of $Z_2$ invariant~\cite{Lang2013,Hung2014}. Furthermore, as to the phase diagram of KMH model, the transition from QSHE to antiferromagnetic Mott insulator has been theoretically predicted to belong to the three-dimensional $XY$ universality class~\cite{Lee2011,Griset2012,Hohenadler2013}, and this continuous transition nature with the universal critical exponents $\beta=0.3486,\nu=0.6717$ is rigorously demonstrated by the finite-size scaling of the $xy$-transverse spin structure factor in the QMC numerical simulations~\cite{Hohenadler2011,Hohenadler2012,Assaad2013a,Hohenadler2014}. Taking into account the rich class of CQPT, it is natural and important to ask whether the CQPT nature is common to different time-reversal symmetric quantum spin Hall systems realized on different lattice geometries.

In this work, we study this interaction-driven transition nature in two representative topological lattice models with time-reversal symmetry through the state-of-the-art density-matrix renormalization group (DMRG) and exact diagonalization (ED) techniques. In Sec.~\ref{model}, we introduce the time-reversal symmetric spinful fermionic Hamiltonian in two typical $\pi$-flux checkerboard and Haldane-honeycomb lattices. In Sec.~\ref{ground}, by tuning the Hubbard repulsion, we demonstrate a CQPT from $Z_2$ QSHE at weak interactions to a trivial Mott antiferromagnetic insulator at strong interactions, with the evidences from Chern number matrix and spin structure factors. In particular, we identify the classification of CQPT is not unique. Specifically, the transition matches with three-dimensional XY universality class in the Haldane-honeycomb lattice, while for the typical $\pi$-flux checkerboard lattice, the transition is possibly in the universality class of the 2D Ising model. Finally, in Sec.~\ref{summary}, we summarize our results and compare the difference between QSHE and integer quantum Hall effect.

\section{Theoretical Models}\label{model}
We consider the spinful fermions in two representative topological lattice models with time-reversal symmetry: (i) the Haldane-honeycomb (HC) lattice~\cite{Wang2011}
\begin{align}
  &H_{HC}^{\uparrow}=-t'\sum_{\langle\langle\rr,\rr'\rangle\rangle}[c_{\rr',\uparrow}^{\dag}c_{\rr,\uparrow}\exp(i\phi_{\rr'\rr})+H.c.]\nonumber\\
  &-t\!\sum_{\langle\rr,\rr'\rangle}\!\!c_{\rr',\uparrow}^{\dag}c_{\rr,\uparrow}
  -t''\!\sum_{\langle\langle\langle\rr,\rr'\rangle\rangle\rangle}\!\!\!\! c_{\rr',\uparrow}^{\dag}c_{\rr,\uparrow}+H.c.,
\end{align}
and (ii) the $\pi$-flux checkerboard (CB) lattice~\cite{Sun2011}
\begin{align}
  &H_{CB}^{\uparrow}=-t\!\sum_{\langle\rr,\rr'\rangle}\!\big[c_{\rr',\uparrow}^{\dag}c_{\rr,\uparrow}\exp(i\phi_{\rr'\rr})+H.c.\big]\nonumber\\
  &-\!\sum_{\langle\langle\rr,\rr'\rangle\rangle}\!\! t_{\rr,\rr'}'c_{\rr',\uparrow}^{\dag}c_{\rr,\uparrow}
  -t''\!\sum_{\langle\langle\langle\rr,\rr'\rangle\rangle\rangle}\!\!\!\! c_{\rr',\uparrow}^{\dag}c_{\rr,\uparrow}+H.c.
\end{align}
Due to time-reversal symmetry, we take $H_{CB}^{\downarrow}=\mathcal{T}H_{CB}^{\uparrow}\mathcal{T}^{-1}$ and $H_{HC}^{\downarrow}=\mathcal{T}H_{HC}^{\uparrow}\mathcal{T}^{-1}$ with $\mathcal{T}$ the time-reversal operation. Here $c_{\rr,\sigma}^{\dag}$ is the particle creation operator of spin $\sigma=\uparrow,\downarrow$ at site $\rr$, $\langle\ldots\rangle$,$\langle\langle\ldots\rangle\rangle$ and $\langle\langle\langle\ldots\rangle\rangle\rangle$ denote the nearest-neighbor, the next-nearest-neighbor, and the next-next-nearest-neighbor pairs of sites, respectively.
Typically, we choose $t''=0,\phi=\pi/2$ for honeycomb lattice which reduces to the famous Kane-Mele (KM) model~\cite{Kane2005a,Kane2005b}, and $t''=0,\phi=\pi/4$ for checkerboard lattice~\cite{Sun2011}. In the flat band limit, we take the parameters $t'=0.6t,t''=-0.58t,\phi=2\pi/5$ for honeycomb lattice and $t'=0.3t,t''=-0.2t,\phi=\pi/4$ for checkerboard lattice.

Taking into account the on-site Hubbard repulsions $V_{int}=U\sum_{\rr}n_{\rr,\uparrow}n_{\rr,\downarrow}$ where $n_{\rr,\sigma}$ is the particle number operator of spin-$\sigma$ at site $\rr$, the model Hamiltonian becomes $H=H_{CB}^{\downarrow}+H_{CB}^{\uparrow}+V_{int}$ ($H=H_{HC}^{\downarrow}+H_{HC}^{\uparrow}+V_{int}$). In the following we explore the many-body ground state of $H$ at half-filling $N_{\uparrow}/N_s=N_{\downarrow}/N_s=1/2$ in a finite system of $N_x\times N_y$ unit cells (the total number of sites is $N_s=2\times N_x\times N_y$) with particle conservation $U(1)\times U(1)$-symmetry. In the ED study, with the translational symmetry, the energy states are labeled by the total momentum $K=(K_x,K_y)$ in units of $(2\pi/N_x,2\pi/N_y)$ in the Brillouin zone. For larger systems we exploit DMRG on the cylindrical geometry, and keep the number of state basis up to 3000 to obtain accurate results.

\section{Interaction-driven phase transitions}\label{ground}

In this section, we present the numerical analysis of the interaction-driven phase transition from two-component QSHE to antiferromagnetism at half-filling. The two-component QSHE can be identified by the Chern number matrix with featureless spin structure factors, and the corresponding charge (spin) pumpings are complementary to and consistent with the Chern number matrix.

\subsection{ED analysis}

\begin{figure}[t]
  \includegraphics[height=2.25in,width=3.4in]{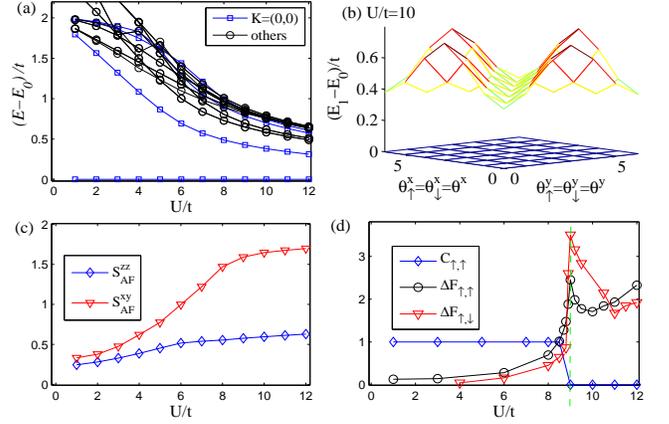}
  \caption{\label{energyhc}(Color online) Numerical ED results for two-component fermions at half-filling $N_s=2\times2\times4=16,N_{\uparrow}=N_{\downarrow}=8$ in the Haldane-honeycomb lattice with $t'=0.3t,t''=0,\phi=\pi/2$. (a) The low energy spectrum as a function of onsite repulsion $U$. (b) The energy spectrum gap for the lowest two energy states in the whole parameter plane $(\theta^{x}_{\uparrow}=\theta^{x}_{\downarrow},\theta^{y}_{\uparrow}=\theta^{y}_{\downarrow})$, keeping $Z_2$ symmetry. (c) The antiferromagnetic spin structure factors $S_{AF}^{zz},S_{AF}^{xy}$ of the ground state as a function of $U$. (d) The topological transition signature of the ground state obtained from its many-body Chern number $C_{\uparrow,\uparrow}$ and the standard deviations of Berry curvature as a function of $U$.
  %(e-g) The spectra flux for the lowest two energies in the whole parameter plane $(\theta_{x}^{\uparrow},\theta_{y}^{\uparrow})$ with $\theta_{x}^{\downarrow}=\theta_{y}^{\downarrow}=0$ at different repulsion $U<U_c,U=U_c,U>U_c$ respectively.
  }
\end{figure}

We first present an ED study of the ground state properties for HC lattice with two different lattice sizes $N_s=16,12$.
In Fig.~\ref{energyhc}(a), we plot the low energy evolution as a function of on-site repulsion $U$.
For weak interactions, there always exists a stable unique ground state at $K=(0,0)$ with a large gap separated from higher levels. By tuning $U$ from weak to strong, there is not any level crossing between the ground state and high-level excited states. Also this ground state does not undergo the level crossing with excited levels in the $Z_2$-symmetric parameter plane $(\theta^{x}_{\uparrow}=\theta^{x}_{\downarrow},\theta^{y}_{\uparrow}=\theta^{y}_{\downarrow})$ as indicated in Fig.~\ref{energyhc}(b), signaling a continuous phase transition nature with $Z_2$ symmetry. (Here $\theta_{\sigma}^{\alpha}$ is the twisted angle for spin-$\sigma$ particles in the $\alpha$-direction, which shifts the particle crystal momentum $\kk_{\alpha}\rightarrow\kk_{\alpha}+\theta_{\sigma}^{\alpha}/N_{\alpha}$; see the definition below). We emphasize that to fully establish the continuous ground energy evolution without level crossing  we need to perform  a scaling of the system size results, which is beyond our current ED limit. Instead, we will demonstrate its continuous transition for large system sizes from the DMRG calculation of ground state wavefunction fidelity and antiferromagnetic order parameters, as shown in Sec.~\ref{dmrg}.
%Although a detailed finite-size scaling is not available in ED calculation, these observations clearly indicate that the first-order transition can be excluded.

Alternatively, the topological index obtained in ED calculation can help us locate the
phase transition boundary. The topological nature of quantum spin-Hall state is characterized by the Chern number matrix by introducing twisted boundary conditions~\cite{Sheng2003,Sheng2006} $\psi(\cdots,\rr_{\sigma}^{i}+N_{\alpha},\cdots)=\psi(\cdots,\rr_{\sigma}^{i},\cdots)\exp(i\theta_{\sigma}^{\alpha})$. The system is periodic when one flux quantum $\theta_{\sigma}^{\alpha}=0\rightarrow2\pi$ is inserted. Meanwhile, the many-body Chern number of the ground state wavefunction $\psi$ is defined as $C_{\sigma,\sigma'}=\frac{1}{2\pi}\int d\theta_{\sigma}^{x}d\theta_{\sigma'}^{y}F_{\sigma,\sigma'}^{xy}$ with the Berry curvature
\begin{align}
  F_{\sigma,\sigma'}^{xy}=\mathbf{Im}\left(\langle{\frac{\partial\psi}{\partial\theta_{\sigma}^x}}|{\frac{\partial\psi}{\partial\theta_{\sigma'}^y}}\rangle
-\langle{\frac{\partial\psi}{\partial\theta_{\sigma'}^y}}|{\frac{\partial\psi}{\partial\theta_{\sigma}^x}}\rangle\right).\nonumber
\end{align}
Due to time-reversal symmetry, for any ground state and interaction, one has the antisymmetric properties $C_{\uparrow,\downarrow}=-C_{\downarrow,\uparrow},C_{\uparrow,\uparrow}=-C_{\downarrow,\downarrow}$ in the spanned Hilbert space, and the total Chern number related to the charge Hall conductance is equal to zero $C_q=\sum_{\sigma,\sigma'}C_{\sigma,\sigma'}=0$ for any interaction strength. Therefore we always have an antisymmetric $C$-matrix~\cite{Sheng2006}
\begin{align}
  \mathbf{C}=\begin{pmatrix}
C_{\uparrow,\uparrow} & C_{\uparrow,\downarrow}\\
C_{\downarrow,\uparrow} & C_{\downarrow,\downarrow}\\
\end{pmatrix}
\end{align}
For decoupled QSHE at weak interactions, we obtain $C_{\uparrow,\uparrow}=1$, and $C_{\uparrow,\downarrow}=0$. However, for strong interactions, the off-diagonal element $C_{\uparrow,\downarrow}$ related to the drag Hall conductance arising from interspecies correlation may be nonzero for two-component quantum Hall effects~\cite{Sheng2003,Sheng2005,Zeng2017,Nakagawa2017}.

\begin{figure}[t]
  \includegraphics[height=1.4in,width=3.4in]{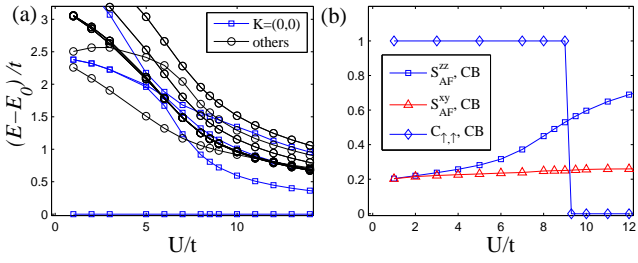}
  \caption{\label{energycb}(Color online) Numerical ED results for two-component fermions at half-filling $N_s=2\times2\times4=16,N_{\uparrow}=N_{\downarrow}=8$ in the $\pi$-flux checkerboard lattice with $t'=0.3t,t''=0,\phi=\pi/4$. (a) The low energy spectrum as a function of onsite repulsion $U$. (b) The antiferromagnetic spin structure factors $S_{AF}^{zz},S_{AF}^{xy}$ of the lowest ground state and its many-body Chern number $C_{\uparrow,\uparrow}$ as a function of $U$. }
\end{figure}

\begin{figure}[t]
  \includegraphics[height=1.4in,width=3.4in]{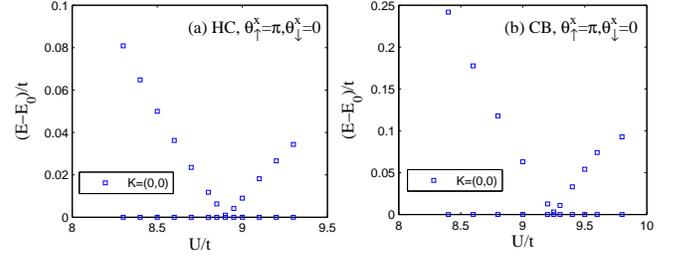}
  \caption{\label{level}(Color online) Numerical ED results for the lowest energy levels at half-filling $N_s=16,N_{\uparrow}=N_{\downarrow}=8$ as a function of Hubbard repulsion under the insertion of half flux quantum $\theta^{x}_{\uparrow}=\pi,\theta^x_{\downarrow}=0$ for (a) the Haldane-honeycomb lattice and (b) the $\pi$-flux checkerboard lattice, respectively. The parameters $t'=0.3t,t''=0$.}
\end{figure}

To clarify the interaction-driven topological transition, we calculate the evolution of $C_{\uparrow,\uparrow}$ as a function of $U$. In Fig.~\ref{energyhc}(d), $C_{\uparrow,\uparrow}$ experiences a fast drop as the interaction $U$ increases across the critical threshold $U_c$, where the distribution of Berry curvature exhibits a singular behavior, signalling the topological phase transition of a many-body system~\cite{Carollo2005,Zhu2006}. As a quantitative measure of the fluctuation of the Berry curvature, we take $\Delta F_{\sigma,\sigma'}=\sqrt{\int d\theta_{\sigma}^{x}d\theta_{\sigma'}^{y}[F_{\sigma,\sigma'}^{xy}-\overline{F}]^2}$ where $\overline{F}$ is the average value. Both $\Delta F_{\uparrow,\uparrow}$ and $\Delta F_{\uparrow,\downarrow}$ show a peak at the critical point where topological invariant changes, resulting from the energy level crossing at $(\theta^{x}_{\uparrow},\theta^{y}_{\uparrow})=(\pi,0)$ (see Fig.~\ref{level} for details). Physically, the sudden jump of $C_{\uparrow,\uparrow}$ and the singularity of $\Delta F_{\uparrow,\uparrow}$ mark the quantum phase transition from QSHE to Mott insulator, while the latter is characterized by gapless spin excitations as shown in Fig.~\ref{dmrghc}(d).

In addition, to get a picture about the Mott insulator in the strongly large-$U$ limit, we calculate the antiferromagnetic spin structure factors
\begin{align}
  S_{AF}^{zz}=\sum_{\alpha,\beta}[S_{AF}^{zz}]^{\alpha,\beta}\label{zafm}
\end{align}
and
\begin{align}
  S_{AF}^{xy}=\sum_{\alpha,\beta}[S_{AF}^{xy}]^{\alpha,\beta},\label{xafm}
\end{align}
with the inner functions defined by
\begin{align}
   [S_{AF}^{zz}]^{\alpha,\beta}&=\frac{1}{N_s}\sum_{i,j}(-1)^{\alpha}(-1)^{\beta}\langle S_{i\alpha}^zS_{j\beta}^z\rangle,\label{szz}\\
   [S_{AF}^{xy}]^{\alpha,\beta}&=\frac{1}{N_s}\sum_{i,j}(-1)^{\alpha}(-1)^{\beta}\langle S_{i\alpha}^{+}S_{j\beta}^{-}+S_{i\alpha}^{-}S_{j\beta}^{+}\rangle,\label{sxy}
\end{align}
where $i,j$ denote unit cells and $\alpha,\beta\in\{A,B\}$ are sublattice indices, $(-1)^{\alpha}=1 (-1)$ for $\alpha=A (B)$. As indicated in Fig.~\ref{energyhc}(c), both $S_{AF}^{zz}$ and $S_{AF}^{xy}$ undergo a smooth evolution, implying a continuous transition. In the Mott regime for Haldane-honeycomb lattice, the transverse $xy$-antiferromagnetism $S_{AF}^{xy}$ dominates.

Similar results have also been obtained for the $\pi$-flux CB lattice, except that the dominant antiferromagnetic order is aligned along the $z$-direction given by $S_{AF}^{zz}$ in the Mott regime. As we will show below, the different nature of antiferromagnetic orders leads to distinct finite-size scaling behaviors of CQPT, depending on the lattice details. As shown in Fig.~\ref{energycb} for $\pi$-flux checkerboard model, the ground energy level is continuously connected between quantum spin-hall state at weak interactions and a trivial Mott antiferromagnetism at strong interactions, as the onsite repulsion is changed. However, in the Mott regime, the dominant antiferromagnetic order is aligned along the $z$-direction given by $S_{AF}^{zz}$, instead of the transverse antiferromagnetism $S_{AF}^{xy}$ in the $xy$-plane. Similarly, the second-order transition from QSHE to Ising antiferromagnetism is also predicted in the correlated Bernevig-Hughes-Zhang model, using dynamical mean field theory~\cite{Hohenadler2013,Yoshida2012,Miyakoshi2013}.

However, under the insertion of flux quantum $\theta^{x}_{\uparrow}=\pi,\theta^x_{\downarrow}=0$ which breaks the $Z_2$ symmetry between spin-up and spin-down particles, the lowest two energy levels indeed cross with each other at the critical point where the Berry curvature becomes singular and topological invariant changes, as indicated in Figs.~\ref{level}(a) and~\ref{level}(b).

\subsection{DMRG results}\label{dmrg}

\begin{figure}[t]
  \includegraphics[height=2.6in,width=3.3in]{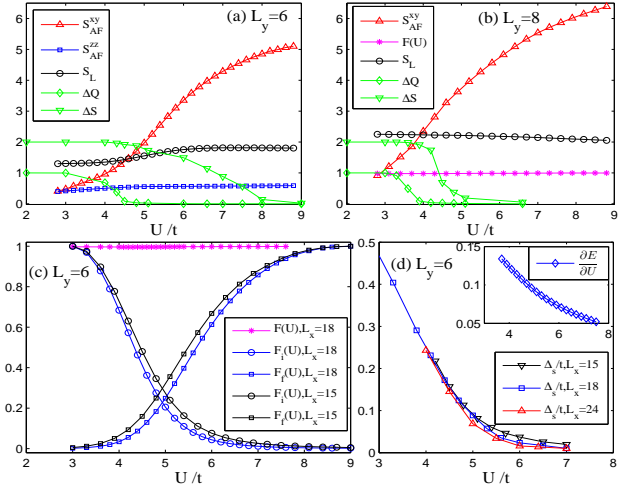}
  \caption{\label{dmrghc}(Color online) Numerical DMRG results on a cylinder Haldane-honeycomb lattice with width $L_y=2N_y$ and length $L_x=N_x=18$ at half-filling. The evolutions of the entanglement entropy $S_{L}$, charge and spin pumpings $\Delta Q,\Delta S$, spin structure factors $S_{AF}^{zz},S_{AF}^{xy}$ as a function of $U$ on a cylinder with width (a) $L_y=6$ and (b) $L_y=8$, respectively; (c) The absolute wavefunction overlap $F(U)=|\langle\psi(U)|\psi(U+\delta U)\rangle|$, $F_i(U)=|\langle\psi(U=3t)|\psi(U)\rangle|$, and $F_f(U)=|\langle\psi(U=9t)|\psi(U)\rangle|$ with different cylinder lengths $L_x=N_x=18,15$ respectively; (d) The spin excitation gap $\Delta_s$ as a function of $U$ for different lengths. The inset shows the ground state energy derivative $\partial E/\partial U$ per site. The smooth transition is characterized by the continuous behavior of these physical quantities. There are no signs of a first-order transition. The parameters $t'=0.3t,t''=0,\phi=\pi/2$.}
\end{figure}

To further verify the continuous interaction-driven transition, we exploit an unbiased DMRG approach for larger system sizes, using a cylindrical geometry up to a maximum width $L_y=8$ ($N_y=4$). As shown in Fig.~\ref{dmrghc} for HC lattice, we measure three different quantities as a function of $U$: the ground state wavefunction overlap $F(U)=|\langle\psi(U)|\psi(U+\delta U)\rangle|$ ($\delta U$ is as small as $0.1t$), the ground state entanglement entropy $S_L$, and ground state energy derivative. We also check the overlaps $F_i(U)=|\langle\psi(U=3t)|\psi(U)\rangle|$ between the ground state $\psi(U)$ and QSHE at $U=3t$, and $F_f(U)=|\langle\psi(U=9t)|\psi(U)\rangle|$ between the ground state $\psi(U)$ and AFM at $U=9t$. All the physical order parameters exhibit continuous evolutions from weak interactions to strong interactions, such that we can exclude the possibility of a first-order phase transition. The spin excitation gap $\Delta_s=E_0(S^z=1)-E_0(S^z=0)$ would tend to diminish continuously in the Mott regime.

\begin{figure}[t]
  \includegraphics[height=1.65in,width=3.4in]{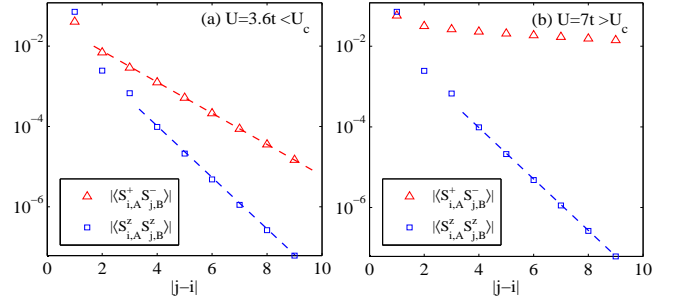}
  \caption{\label{corre}(Color online) Numerical DMRG results of the long-range antiferromagnetic spin correlation functions $|\langle S_{i,A}^{+}S_{j,B}^{-}\rangle|$,$|\langle S_{i,A}^zS_{j,B}^z\rangle|$ as a distance $|j-i|$ in the $x$-direction between sublattice A and sublattice B on a cylinder Haldane-honeycomb lattice with finite width $L_y=2N_y=6$ and fixed length $L_x=N_x=18$ at half-filling for different Hubbard repulsions: (a) $U=3.6t<U_c$ and (b) $U=7.0t>U_c$, respectively. The blue/red dashed lines are the exponential fit to the decaying behaviors of these correlation functions. The parameters $t'=0.3t,t''=0,\phi=\pi/2$.}
\end{figure}

Second, we characterize the topological nature of the ground state from its topological charge pumping by inserting one flux quantum $\theta_{\uparrow}^{y}=\theta,\theta_{\downarrow}^{y}=0$ from $\theta=0$ to $\theta=2\pi$ on cylinder systems based on the newly developed adiabatic DMRG~\cite{Gong2014} in connection to the quantized Hall conductance. The net transfer of the total charge from the right side to the left side is encoded by the expectation value $Q(\theta)=N_{\uparrow}^{L}+N_{\downarrow}^{L}=tr[\widehat{\rho}_L(\theta)\widehat{Q}]$. Here we partition the lattice system  on the cylinder along the $y$-direction into two halves with equal lattice sites. $N_{\sigma}^{L}$ is the particle number of spin-$\sigma$ in the left cylinder part, and $\widehat{\rho}_L$ the reduced density matrix of the corresponding left part~\cite{Zaletel2014}. Under the inserting of the flux  $\theta_{\uparrow}^{y}=\theta,\theta_{\downarrow}^{y}=0$ in the $y$-direction, the change of $N_{\uparrow}^{L}+N_{\downarrow}^{L}$ indicates the transverse charge transfer from the right side to the left side in the $x$-direction, induced by both diagonal Hall conductance $C_{\uparrow,\uparrow}$ and drag Hall conductance $C_{\downarrow,\uparrow}$. From the Chern number matrix of two-component quantum Hall effects, in each cycle we obtain~\cite{Zeng2017}
\begin{align}
  \Delta Q=Q(2\pi)-Q(0)=C_{\uparrow,\uparrow}+C_{\downarrow,\uparrow}.
\end{align}
In order to quantify the spin-Hall conductance, we also calculate the spin pumping by inserting one flux quantum $\theta_{\uparrow}^{y}=\theta_{\downarrow}^{y}=\theta$ from $\theta=0$ to $\theta=2\pi$ in the $y$-direction, and define the $Z_2$ spin transfer $\Delta S$ from the right side to the left side in the $x$-direction by the physical quantity $S(\theta)=N_{\uparrow}^{L}-N_{\downarrow}^{L}=tr[\widehat{\rho}_L(\theta)\widehat{S}]$ in analogy to the charge transfer. Similarly, we obtain~\cite{Zeng2017}
\begin{align}
  \Delta S=S(2\pi)-S(0)=C_{\uparrow,\uparrow}-C_{\downarrow,\uparrow}+C_{\uparrow,\downarrow}-C_{\downarrow,\downarrow}.
\end{align}
For each flux cycle, we obtain both $\Delta Q\simeq1$ and $\Delta S\simeq2$ for the QSHE in the weakly interacting regime. However in the strongly interacting regime $\Delta Q\simeq0,\Delta S\simeq0$. The change of the charge pumping is shown in Fig.~\ref{dmrghc}(a), where the critical $U_c\simeq4.8t$, while the $Z_2$ spin pumping persists a finite value deviating from the integer quantized value $2$ up to $U_{c}'>U_c$. However, with increasing $L_y=6$ to $L_y=8$, we find that the difference between $U_c$ and $U_c'$ becomes substantially reduced as shown in Figs.~\ref{dmrghc}(a-b), which may be consistent with a direct transition from the QSHE to the Mott insulator.

\begin{figure}[t]
  \includegraphics[height=1.5in,width=3.4in]{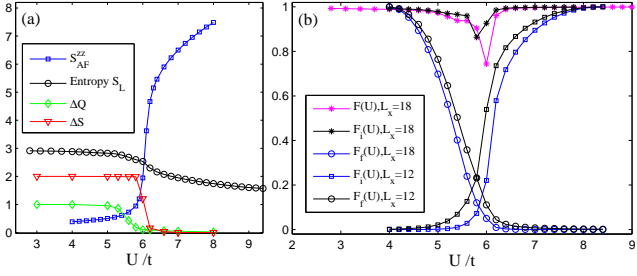}
  \caption{\label{dmrgcb}(Color online) Numerical DMRG results on a cylinder $\pi$-flux checkerboard lattice with width $L_y=2N_y=6$ at half-filling with parameters $t'=0.3t,t''=0$. (a) The evolutions of the entanglement entropy $S_{L}$, charge and spin pumpings $\Delta Q,\Delta S$, spin structure factors $S_{AF}^{zz}$ as a function of $U$ on a cylinder with length $L_x=N_x=18$. (b) The absolute wavefunction overlap $F(U)=|\langle\psi(U)|\psi(U+\delta U)\rangle|$, $F_i(U)=|\langle\psi(U=3t)|\psi(U)\rangle|$, and $F_f(U)=|\langle\psi(U=9t)|\psi(U)\rangle|$ with different cylinder lengths $L_x=N_x=18,12$ respectively.}
\end{figure}

Third, we measure the antiferromagnetic order from spin structure factors $S_{AF}^{zz},S_{AF}^{xy}$. Both $S_{AF}^{zz},S_{AF}^{xy}$ exhibit a continuous evolution near the critical point for different system sizes as shown in Figs.~\ref{dmrghc}(a) and~\ref{dmrghc}(b), similar to our ED analysis. In the thermodynamic limit, $S_{AF}^{zz}$ should be vanishingly small in the strong interacting limit $U\gg t$. In Figs.~\ref{corre}(a) and~\ref{corre}(b), our DMRG results show that for $U<U_c$, both of the antiferromagnetic spin correlations $\langle S_{i,A}^{+}S_{j,B}^{-}\rangle,\langle S_{i,A}^zS_{j,B}^z\rangle$ decay exponentially as the distance $|j-i|$, while for $U>U_c$, only the longitudinal long-range order parameters $\langle S_{i,A}^zS_{j,B}^z\rangle$ decays exponentially as the distance $|j-i|$, but the transverse long-range order parameters $\langle S_{i,A}^{+}S_{j,B}^{-}+S_{i,A}^{-}S_{j,B}^{+}\rangle$ maintain to be a robust finite value of the order 0.01, which determines the square of transverse XY spontaneous magnetization $m_{xy}^2=\lim_{|j-i|\rightarrow\infty}|\langle S_{i,A}^{+}S_{j,B}^{-}+S_{i,A}^{-}S_{j,B}^{+}\rangle|$. For $|j-i|>6$, $\langle S_{i\alpha}^zS_{j\beta}^z\rangle$ becomes already smaller than $10^{-6}$, These are in good agreement with the physical picture proposed in Refs.~\cite{Rachel2010,Hohenadler2012,Assaad2013a,Hohenadler2014}. For our study, a very small value of spin structure factor $S_{AF}^{zz}$ in both ED and DMRG, is likely due to the finite width effects in the $y$-direction. Figures.~\ref{dmrgcb}(a) and~\ref{dmrgcb}(b) show the continuous phase transition through the tunable repulsion $U$ on a cylinder $\pi$-flux checkerboard lattice for large system sizes. The topological phase transition is characterized by the charge and spin pumpings when inserting one flux quantum. All the physical order parameters like spin structure factor, entanglement entropy and the wavefunction fidelity exhibit a continuous evolution from weak interactions to strong interactions.

As shown in Figs.~\ref{dmrghc} and~\ref{dmrgcb}, for HC lattice, only $S_{AF}^{xy}$ shows a rapid increase, signaling an antiferromagnetic order in the transverse $xy$-plane for $U>U_c$; In contrast for $\pi$-flux CB lattice, only $S_{AF}^{zz}$ shows a rapid increase near the critical point, signaling an antiferromagnetic order in the longitudinal $z$-direction for $U>U_c$. To understand this, let us consider the antiferromagnetic long-range-ordered phase in the strongly large-$U$ limit. For $U\gg t$, similar to the usual Hubbard model, we expand the Hamiltonian in powers of $t/U$ up to the second order, and arrive at the effective spin models $J\sum_{\langle\rr,\rr'\rangle}[S_{\rr}^{z}S_{\rr'}^{z}+i(S_{\rr}^{+}S_{\rr'}^{-}-S_{\rr}^{-}S_{\rr'}^{+})/2]
+J'\!\sum_{\langle\langle\rr,\rr'\rangle\rangle}\mathbf{S}_{\rr}\cdot\mathbf{S}_{\rr'}
+J''\!\sum_{\langle\langle\langle\rr,\rr'\rangle\rangle\rangle}\mathbf{S}_{\rr}\cdot\mathbf{S}_{\rr'}$ for $\pi$-flux CB lattice and $J\sum_{\langle\rr,\rr'\rangle}\mathbf{S}_{\rr}\cdot\mathbf{S}_{\rr'}
+J'\!\sum_{\langle\langle\rr,\rr'\rangle\rangle}[S_{\rr}^{z}S_{\rr'}^{z}+(e^{2i\phi}S_{\rr}^{+}S_{\rr'}^{-}+e^{-2i\phi}S_{\rr}^{-}S_{\rr'}^{+})/2]
+J''\!\sum_{\langle\langle\langle\rr,\rr'\rangle\rangle\rangle}\mathbf{S}_{\rr}\cdot\mathbf{S}_{\rr'}$ for Haldane HC lattice, where $J=4t^2/U, J'=4(t')^2/U,J''=4(t'')^2/U$~(see also the related effective spin Hamiltonian for HC lattice in Refs.~\cite{Rachel2010,Reuther2012}).

\begin{figure}[t]
  \includegraphics[height=2.4in,width=3.4in]{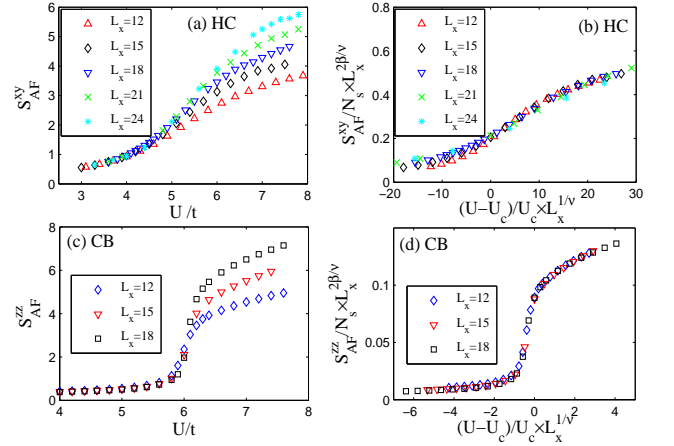}
  \caption{\label{scalexy}(Color online) Numerical DMRG results for the spin structure factor on a cylinder with width $L_y=2N_y=6$ at half-filling with parameters $t''=0$. (a-b) Finite-size dependence of the structure factor $S_{AF}^{xy}$ as a function of $U$ for HC lattice with critical exponents $\beta\simeq0.31,\nu\simeq0.7$ at $t'=0.3t$. (c-d) Finite-size scaling of the structure factor $S_{AF}^{zz}$ as a function of $U$ for CB lattice with critical exponents $\beta\simeq1/8,\nu\simeq1.0$ at $t'=0.3t$.}
\end{figure}

When $t''=0$, for $\pi$-flux CB lattice the nearest-neighbor term is an Ising exchange, while the next-nearest-neighbor term is an isotropic antiferromagnetic Heisenberg exchange. However for Haldane HC lattice the nearest-neighbor term is an isotropic antiferromagnetic Heisenberg exchange, while the next-nearest-neighbor term is antiferromagnetic in the longitudinal direction but ferromagnetic in the transverse direction when $\phi$ is close to $\pi/2$. In our typical parameters $J'/J\lesssim0.3$ which is away from possible spin liquid regime~\cite{Gong2014a,Gong2013}, combining all the exchange terms, we expect an antiferromagnetic order in the $z$-direction for $\pi$-flux checkerboard lattice, but in the $xy$-plane for Haldane-honeycomb lattice, due to the next-nearest-neighbor frustration term. As a result, the scaling behavior around the critical point is different for HC and CB lattices.

\begin{figure}[t]
  \includegraphics[height=1.35in,width=3.3in]{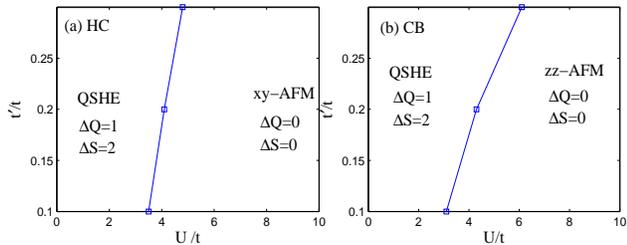}
  \caption{\label{phase}(Color online) Numerical DMRG results of the phase diagram for (a) Haldane-honeycomb lattice and (b) $\pi$-flux checkerboard lattice models on a cylinder with finite width $L_y=2N_y=6$ and fixed length $L_x=N_x=18$ at half-filling for $t''=0$.}
\end{figure}

Due to the numerical difficulty of well-controlled DMRG convergence for two-component particles on cylinder width $L_y>8 (N_y>4)$, we cannot perform a finite-size scaling in the $y$-direction, and therefore focus on the quasi-one dimensional scaling of the cylinder length $L_x$. This is different from QMC methods where the finite-size scaling is done at the same time in both $x,y$-directions. Despite its limitation, we show that it still sheds some light into the critical scaling exponents. For HC lattice, in Figs.~\ref{scalexy}(a) and~\ref{scalexy}(b), a finite size scaling of $S_{AF}^{xy}$ by using the scaling function  $S_{AF}^{xy}/N_s\propto L_{x}^{-2\beta/\nu}f(L_{x}^{1/\nu}(U-U_c))$ gives the critical exponents $\beta=0.31,\nu=0.70$. For QMC simulations in Refs.~\cite{Hohenadler2012,Assaad2013a,Assaad2013b}, they extract the exponents $\beta=0.3486,\nu=0.6717$ in fully agreement with those of 3D XY model. In comparison, we can see that our DMRG results are in reasonable agreement with 3D XY universality class. While for CB lattice in Figs.~\ref{scalexy}(c) and~\ref{scalexy}(d), $S_{AF}^{zz}$ from different sizes can merge together by using the scaling function  $S_{AF}^{zz}/N_s\propto L_{x}^{-2\beta/\nu}f(L_{x}^{1/\nu}(U-U_c))$
using the critical exponents $\beta=1/8,\nu=1$, which indicates that the phase transition falls into the 2D Ising universality class~\cite{Pelissettoa2002,Moukouri2012}. When $U$ approaches a critical value $U_c$, $F(U)$ shows a small bump, implying a peak of the fidelity susceptibility $\chi_F=(1-F(U))/(\delta U)^2$ which is a signature of phase transition~\cite{Gu2010,Saadatmand2017}. We obtain a similar picture for $\chi_F\propto L_{x}^{2/\nu}f(L_{x}^{1/\nu}(U-U_c))$. Thus we conjecture that this phase transition maybe belong to the 2D Ising universality class, which is different from that in HC lattice, although a stronger evidence of finite-size scaling in cylinder width is necessary.

\begin{figure}[t]
  \includegraphics[height=1.5in,width=3.2in]{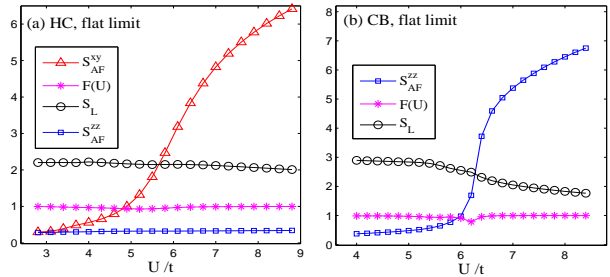}
  \caption{\label{flat}(Color online) Numerical DMRG results on a cylinder lattice in the flat band limit with width $L_y=2N_y$ and length $L_x=N_x=18$ at half-filling. The evolutions of the absolute wavefunction overlap $F(U)$, spin structure factors $S_{AF}^{zz},S_{AF}^{xy}$, and entanglement entropy $S_{L}$ are shown for (a) the Haldane-honeycomb lattice with  parameters $t'=0.6t,t''=0.58t,\phi=2\pi/5$ and (b) the checkerboard lattice with parameters $t'=0.3t,t''=-0.2t,\phi=\pi/4$, respectively. The smooth transition is characterized by the continuous behavior of these physical quantities.}
\end{figure}

Finally, we present our DMRG results of the phase diagram in the parameter plane $(U,t')$ without $t''$, as indicated in Figs.~\ref{phase}(a) and~\ref{phase}(b). First of all, we identify a CQPT separating the QSHE from the antiferromagnetic ground state on both HC and CB lattices, without the evidence of intermediate phase in between. The apparent non-vanishing spin pump is due to the fluctuating off-diagonal Berry curvature driven by interspecies correlation, as also identified from ED analysis in Fig.~\ref{energyhc}(d). Here we do not consider the situation $t'\rightarrow0$ where the system is a gapless Dirac semimetal for both models, and the transition from such a Dirac semimetal to AFM has been claimed to be of Gross-Neveu universality class in several QMC simulations~\cite{Assaad2013a,Assaad2013b}, which we leave  for  future study.
Moreover, by including the next-next-nearest-neighbor hopping in the flat band limit, we obtain the similar physical picture of a continuous phase transition. As shown in Fig.~\ref{flat}(a) and~\ref{flat}(b), for both Haldane-honeycomb and $\pi$-flux checkerboard, the quantum phase transition is continuous, identified from three physical quantities including the absolute wavefunction overlap $F(U)$, spin structure factors $S_{AF}^{zz},S_{AF}^{xy}$, and entanglement entropy $S_{L}$.

\section{Summary and Discussions}\label{summary}
In summary, using both ED and DMRG calculations, we have demonstrated a continuous phase transition from a quantum spin-Hall state to an antiferromagnetic Mott insulator driven by onsite Hubbard repulsion at half-filling, which is characterized by the continuous evolutions of the physical quantities, including the wave function fidelity, spin structure factors, entanglement entropy. The topological transition nature is encoded by the singular behavior of the Berry curvature driven by strong interspecies correlation, but the total charge Hall conductance remains unchanged. In close comparison, for an integer quantum Hall state with a symmetric $C$-matrix $\mathbf{C}=\begin{pmatrix}
1 & 0\\
0 & 1\\
\end{pmatrix}$ in both $\pi$-flux checkerboard and Haldane-honeycomb lattices with broken time-reversal symmetry, recent ED and DMRG studies show that a direct first-order level crossing occurs at one of high-symmetry twisted boundary conditions~\cite{Zeng2017,Vanhala2016} from integer quantum Hall effect (IQHE) to a trivial Mott insulator. Physically,
these two classes of transitions indeed can belong to different classes as the transition between IQHE to a Mott insulator has a quantized jump of charge Chern number between two charge insulators, while the transition between QSHE and a Mott insulator only has a change of spin Chern number from a spin insulator to a gapless spin system.
We believe our current work may provide a new insight into the interaction-driven topological transition nature. As one
intriguing direction of our study, it is interesting and important to investigate the role of broken U(1)-spin symmetry by adding spin-orbit coupling in the transition between QSHE and the magnetic phase, which is very relevant to the transition metal oxide Na$_{2}$IrO$_{3}$ materials~\cite{Shitade2009,Jackeli2009}, and we leave it for a future follow-up project.

\begin{acknowledgements}
W.Z. thanks Kai Sun for stimulating discussions.
This work is supported by National Science Foundation Grants PREM DMR-1205734 (T.S.Z.) and DMR-
1408560 (D.N.S.)
% U.S. Department of Energy Contract No. DE-AC52-06NA25396 and No. DR-20170664PRD1 through the LDRD Program (W.Z.).
The work at Los Alamos was supported by the U.S. DOE Contract No. DE-AC52-06NA25396
through the LDRD Program (W.Z. \& J.-X.Z.),
and supported in part by the Center for Integrated Nanotechnologies,
a U.S. DOE Office of Basic Energy Sciences user facility.
\end{acknowledgements}

\end{document}